\documentclass{article}

\usepackage[numbers,compress,sort]{natbib}

\usepackage[final]{ai4science}

\usepackage[utf8]{inputenc} 
\usepackage[T1]{fontenc}    
\usepackage{hyperref}       
\usepackage{url}            
\usepackage{booktabs}       
\usepackage{amsfonts}       
\usepackage{nicefrac}       
\usepackage{microtype}      
\usepackage{xcolor}         
\usepackage{amsmath}
\usepackage[pdftex]{graphicx}
\usepackage{caption}
\usepackage{subcaption}
\usepackage{mathabx}

\title{Predicting Drug-Drug Interactions using Deep Generative Models on Graphs}

\author{%
Nhat Khang Ngo\thanks{Co-first authors. $\dagger$ Corresponding author.}\\
FPT AI Center \\
Hanoi, Vietnam \\
khangnn3@fsoft.com.vn
\And
Truong Son Hy $^\dagger$ \footnotemark[1] \\
University of California San Diego\\
La Jolla, CA 92093, US \\
tshy@ucsd.edu
\And
Risi Kondor\\
University of Chicago\\
Chicago, IL 60637, US \\
risi@uchicago.edu
}

\begin{document}

\maketitle

\begin{abstract}

Latent representations of drugs and their targets produced by contemporary graph autoencoder-based models have proved useful in predicting many types of node-pair interactions on large networks, including drug-drug, drug-target, and target-target interactions. However, most existing approaches model the node's latent spaces in which node distributions are rigid and disjoint; these limitations hinder the methods from generating new links among pairs of nodes. In this paper, we present the effectiveness of variational graph autoencoders (VGAE) in modeling latent node representations on multimodal networks. Our approach can produce flexible latent spaces for each node type of the multimodal graph; the embeddings are used later for predicting links among node pairs under different edge types. To further enhance the models' performance, we suggest a new method that concatenates Morgan fingerprints, which capture the molecular structures of each drug, with their latent embeddings before preceding them to the decoding stage for link prediction. Our proposed model shows competitive results on two multimodal networks: (1) a multi-graph consisting of drug and protein nodes, and (2) a multi-graph consisting of drug and cell line nodes. Our source code is publicly available at \url{https://github.com/HySonLab/drug-interactions}.
\end{abstract}

\section{Introduction}
Investigations on biomedical multimodal graphs consisting of drugs, proteins, or cells and their interrelations are essential in drug repurposing as they can help discover innovative node-pair interactions, which may be helpful in medical treatments, on these networks. Most approaches to tackling drug repurposing problems estimate drug side effects and drug responses to cell lines to provide important information about the use of those drugs. In particular, polypharmacy is the practice of treating complex diseases or co-existing health conditions by using combinations of multiple medications \cite{bansal2014community}. Drug combinations consisting of many drugs affecting distinct targeted proteins can effectively modulate the process of severe diseases by facilitating the diverse biological process of different proteins \cite{sun2015combining}. Albeit commonly applied, polypharmacy is one of the major underlying issues that cause adverse medical outcomes, also known as side effects caused by drug combinations \cite{10.1093/bioinformatics/bty294}. Understanding drug-drug interactions (DDIs) is pivotal in predicting the potential side effects of multiple co-administered medications. Moreover,  accurate estimation of drug responses to targeted cell lines is another major factor in providing precision medicines to patients. However, it is unfeasible to render clinical testing of all drug combinations due to the tremendous number of relations between drug pairs or drugs and their targets (e.g., proteins). 

Machine learning models have been widely applied to provide tractable solutions in predicting potential drug interactions. Data-driven techniques (e.g., machine learning or deep learning) have proved their capabilities for dealing with complex patterns in data. Such methods have produced remarkable results in various fields, such as computer vision, natural language processing, and audio processing. Therefore, it is reasonable to employ machine learning approaches to investigate the interactions among complex combinations of medications. Previous work on DDIs \cite{gottlieb2012indi, vilar2012drug, cheng2014machine} used various hand-crafted drug features (e.g., chemical properties or Morgan fingerprints) to compute the similarity of each drug pair, and drug-drug interactions can be predicted based on the drug-pair proximity scores. Nevertheless, using fixed drug representations can result in sub-optimal outcomes since they do not sufficiently capture the complex interrelations of drugs.  

Recent approaches rely on graph neural networks (GNNs) \cite{battaglia2018relational} to directly learn drug and protein representations from structural data (i.e. graphs) augmented with additional information for node features. \citet{10.1093/bioinformatics/bty294} proposed a graph autoencoder to predict polypharmacy side effects on a multimodal graph consisting of drug and protein nodes with various edge types. Similarly, \citet{yin2022deepdrug} introduced DeepDrug combining graph convolutional neural networks (GCN) \cite{Kipf:2017tc} and convolutional neural networks (CNN)\cite{lecun1995convolutional} to learn the structural and sequential representations of drugs and proteins. In addition, \citet{wang2022deepdds} suggested using several graph attention (GAT) \cite{velickovic2018graph} layers to learn on hand-crafted features of drugs to predict drug-pair interactions. \citet{nyamabo2021ssi} use multiple GAT layers to compute the drug representations and aggregate them using co-attention to produce final predictions. These mentioned GNN-based models effectively learn the latent representations of drugs and proteins, which can be used in several downstream tasks. 

Predicting drug-drug or drug-target interactions can be regarded as a link prediction task on graphs. We argue that latent spaces produced by the aforementioned approaches are disjoint and not interpretable; hence, they are not capable of generating new links on some graph benchmarks. In this paper, we suggest using deep graph generative models to make the latent spaces continuous and more interpretable; as a result, they are superior to traditional graph autoencoders (GAE) in predicting new links on biomedical multimodal graphs. Instead of yielding deterministic latent representations, variational graph autoencoders (VGAE) \cite{kipf2016variational} use a probabilistic approach to compute the latent variables. In this work, we aim to use VGAE to learn the representations of drugs and proteins on a multimodal graph; then, we predict several node-pair interactions (e.g., drug-pair polypharmacy side effects, drug-cell line response, etc.) using the learned representations. 

Our contributions are two-fold as follows:
\begin{itemize}
	\item We demonstrate that using VGAE can attain superior performance in predicting drug-drug interactions on multimodal networks, outperforming GAE-based architectures and hand-crafted feature based methods.
	\item We leverage drug molecular representations (i.e. Morgan fingerprints) concatenated with drug latent representations before the decoding stage to further enhance the model's performance in predicting drug-pair polypharmacy side effects. 
\end{itemize}

\section{Related work}
Link prediction is one of key problems in graph learning. In link-level tasks, the network consists of nodes and an incomplete set of edges between them, and partial information on the graph are used to infer the missing edges. The task has a wide range of variants and applications such as knowledge graph completion \cite{10.5555/2886521.2886624}, link prediction on citation networks \cite{bojchevski2018deep}, and protein-protein interaction prediction \cite{Zitnik2017}.  

Graph neural networks (GNNs) are deep learning models that learn to generate representations on graph-structured data. Most GNN-based approaches use a message passing paradigm proposed in \cite{4700287} \cite{pmlr-v70-gilmer17a} wherein each node aggregates vectorized information from its neighbors and updates the messages to produce new node representations. \citet{Kipf:2017tc} introduced a scalable method that approximates first-order spectral graph convolutions, which are equivalent to one-hop neighborhood message passing neural networks. Besides, \citet{velickovic2018graph} proposed graph attention networks (GATs) that use soft attention mechanism to weight the messages propagated by the neighbors. \citet{NIPS2017_5dd9db5e} introduced GraphSAGE that efficiently uses node attribute information to generate representations on large graphs in inductive settings.

Deep generative models aim to generate realistic samples, which should satisfy properties in nature. Different from traditional methods that rely on hand-crafted features, data-driven approaches can be categorized into variational autoencoders (VAEs) \cite{kingma2013auto}, generative adversarial networks (GANs) \cite{NIPS2014_5ca3e9b1}, and autoregressive models. In the field of deep generative models on graphs, several works (\cite{pmlr-v70-ingraham17a, NEURIPS2020_41d80bfc} aim at adding stochasticity among latents to make the models more robust to complicated data distributions. \citet{graphvae} proposed a regularized graph VAE model wherein the decoder outputs probabilistic graphs. In addition, \citet{netgan} introduced a stochastic neural networks that are trained by Wassterstain GAN object to learn the distribution of random walks on graphs. For autoregressive methods, \citet{gran} proposed a attention-based GNN network to generate graph nodes and associated edges after many decision steps in a generation process.

The objective of our work is to apply a probabilistic framework in predicting missing links on biomedical multimodal networks. We extend varational graph autonecoders (VGAE) \cite{kipf2016variational} to multimodal networks such that separate posterior latent distributions are computed for each node type after several message passing and aggregation layers among different node and edge types. Moreover, we also incorporate molecular structures into the latent variables of drug nodes to capture more inter-intra information across the medications, resulting in superior performance in drug-pair prediction.

\section{Approach}
\subsection{Problem Setup}
Given a multimodal graph, our task is to predict whether two nodes of the same or different types are connected or not under a specific edge type. This can be regarded as a link-level prediction on graphs. In our work, the graph $G$ consists of two node types (e.g., drug and protein), and each side effect is represented as an edge type. In addition to polypharmacy interactions, $G$ also involves edges representing drug-protein and protein-protein interactions. Formally, let denote $G = (V, E, X)$, where $V = V_d \cup V_p$ is a union of two node sets of different types (i.e. $V_d$ is the set of drug nodes and $V_p$ is the set of protein nodes), $E$ is a set of edges, and $X = X_d \oplus X_p$ is a concatenated matrix denoting the node features of different node types. Each edge in $E$ is a triplet $(v_i, e, v_j)$ in which node $v_i$ interacts with node $v_j$ under a specific edge type $e$. In case of a binary classification setting, we perform negative sampling to generate non-edge labels to make the model robust to negative examples. Thus, the objective is to learn a function $f: E \rightarrow T$, where $f$ predicts the value of a particular triplet $(v_i, e, v_j)$; $T$ can be either $\{0,1\}$ or $\mathbb{R}$.
\subsection{Model Architecture}
\begin{figure}[h]
     \centering
     \begin{subfigure}[b]{0.4\textwidth}
         \centering
         \includegraphics[width =\textwidth]{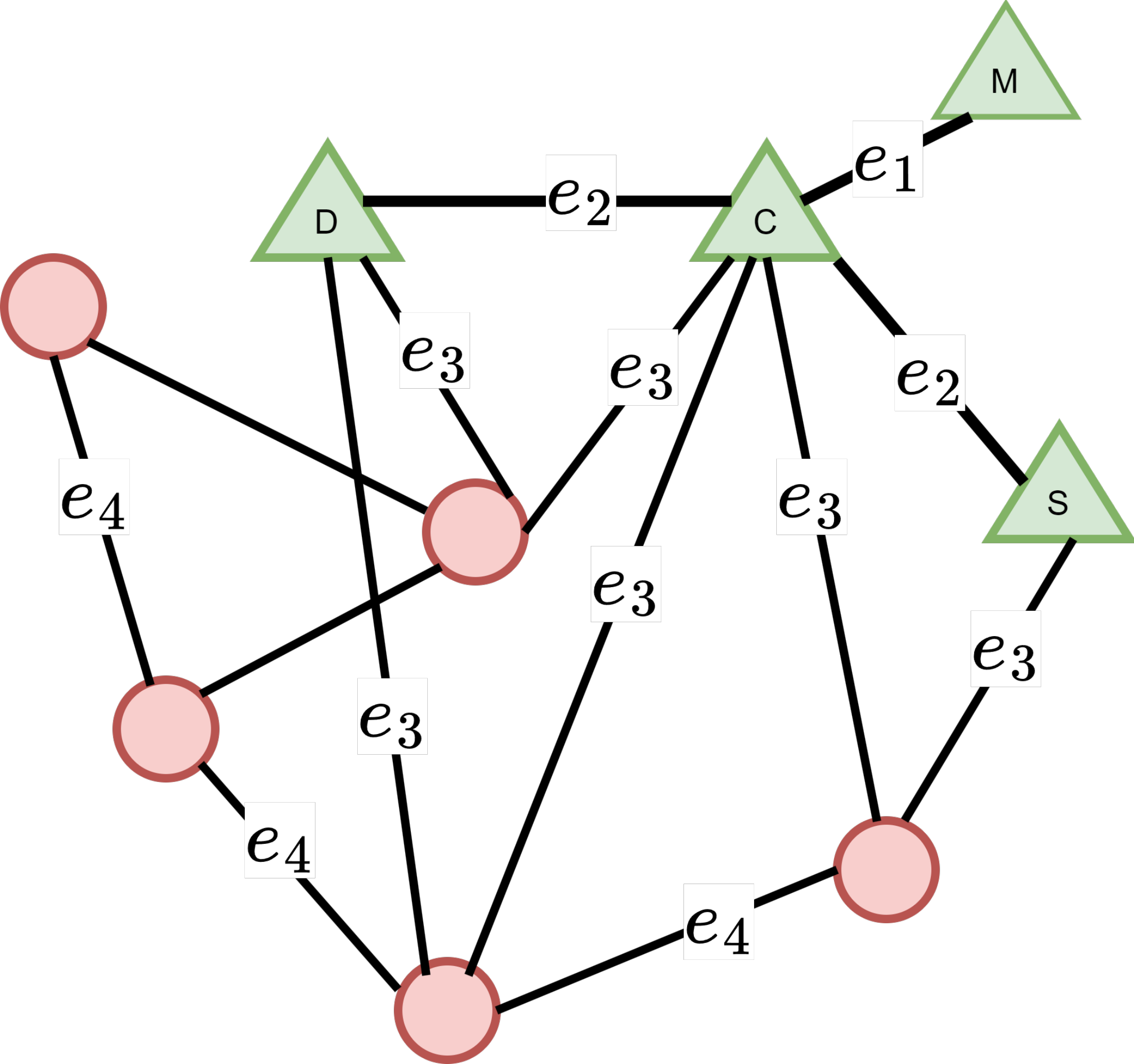}
         \caption{}
         \label{fig:graph}
     \end{subfigure}
     \hfill
     \begin{subfigure}[b]{0.5\textwidth}
         \centering
         \includegraphics[width =\textwidth]{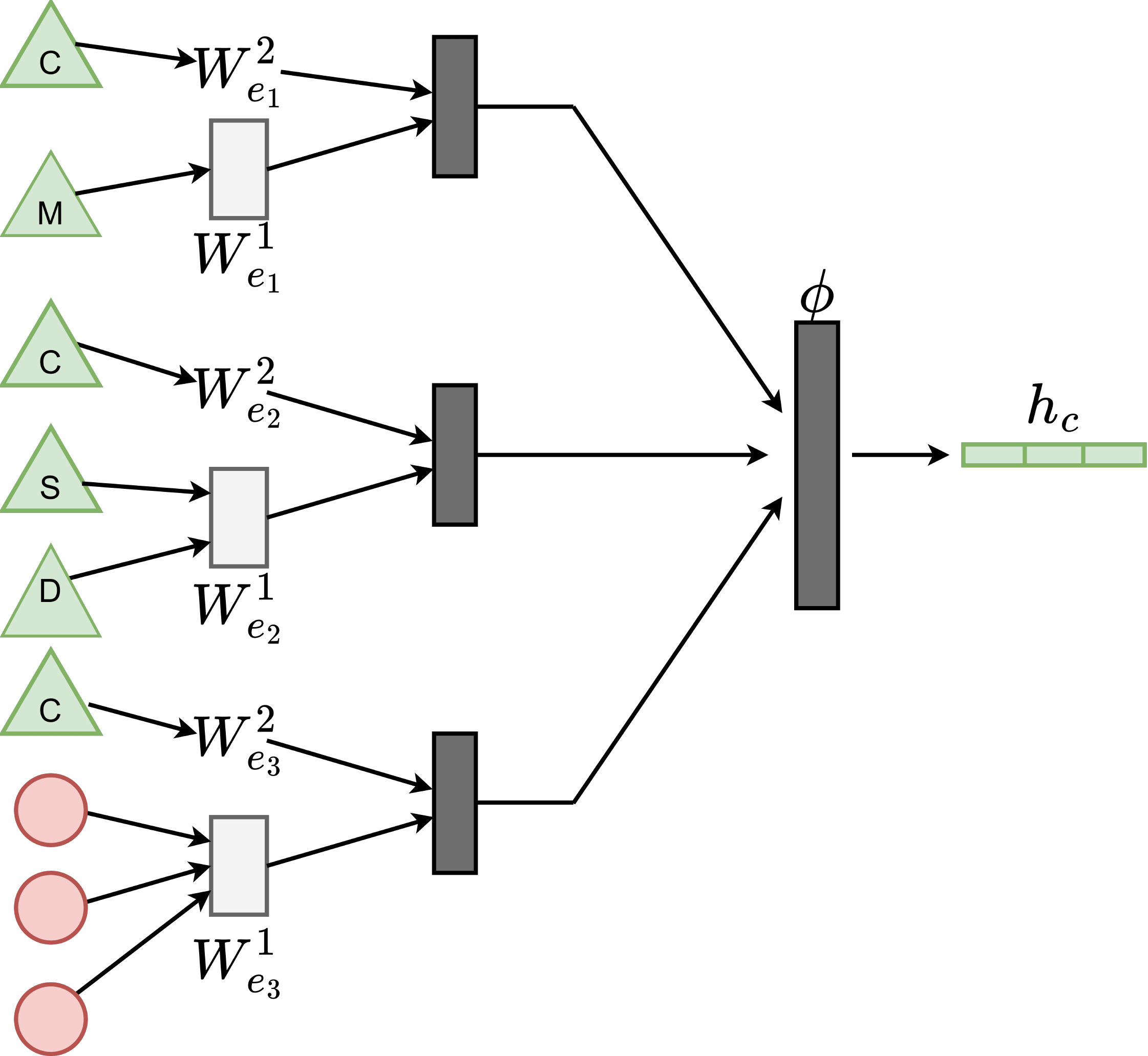}
         \caption{}
         \label{fig:encoder}
     \end{subfigure}
     \vskip 1.2em
        \caption{Overview of a biomedical multimodal graph and one graph convolution layer in our framework. (a) A graph consists of two node types (e.g., red nodes are protein and the green are drug nodes) and five edge types $\{e_i\}$. (b) A layer of the encoder in which the representation of the source node $c$ is computed by aggregating its neighbors' information under different edge types. Black rectangles denote aggregation , whereas white rectangles indicate neural networks that share parameters across the nodes. $\{(W^1_{e_i}, W^2_{e_i})\}$ denote trainable weight matrices of different edge types $\{e_i\}$. $h_c$ can be preceded to either a successive convolution layer or the latent encoder.}
        \label{fig:2}
\end{figure}
\label{sec:sub2}
Our proposed model has three main components: an encoder, a latent encoder, and a decoder. They are described as follows:

\paragraph{Encoder}\label{latent_encoder} a two-layer graph convolutional neural network operating message passing scheme \cite{pmlr-v70-gilmer17a}  on $G$ and producing node embeddings for each node type (e.g., drugs and proteins). Each layer in the network has the following form: 
    \[h_i = \phi \big(\sum_{e}\sum_{j \in \mathcal{N}^i_e \cup \{i\}} W_{e}\frac{1}{\sqrt{c_i c_j}} x_j\big)\]
    where  $\mathcal{N}^i_e$ denotes the neighbor set of node $x_i$ under the edge type $e$.
    $W_{e} \in \mathbb{R} ^ {d_k \times d}$ is a edge-type specific transformation matrices that map $x_i \in \mathbb{R} ^ {d_i}$ and its neighbors $x_j \in \mathbb{R} ^ {d_j}$ into $d_k$-dimensional vector spaces, resulting in $h_i \in \mathbb{R}^{d_k}$. It is worth noting that $d_i$ and $d_j$ are not necessarily the same because $x_i$ and $x_j$ can be nodes of two different node types (i.e. $d_i = d_j$ when $x_i$ and $x_j$ are in the same node type). $c_i$ and $c_j$ indicates the degree node $i$ and $j$ on the network. Also, $\phi$ denotes a nonlinear activation function; we use rectified linear unit (ReLU) \cite{6639346} in our experiments. Figure \ref{fig:2} illustrates an overview of the encoder in our framework.

\begin{figure}[h]
    \centering
    \includegraphics[scale=0.5]{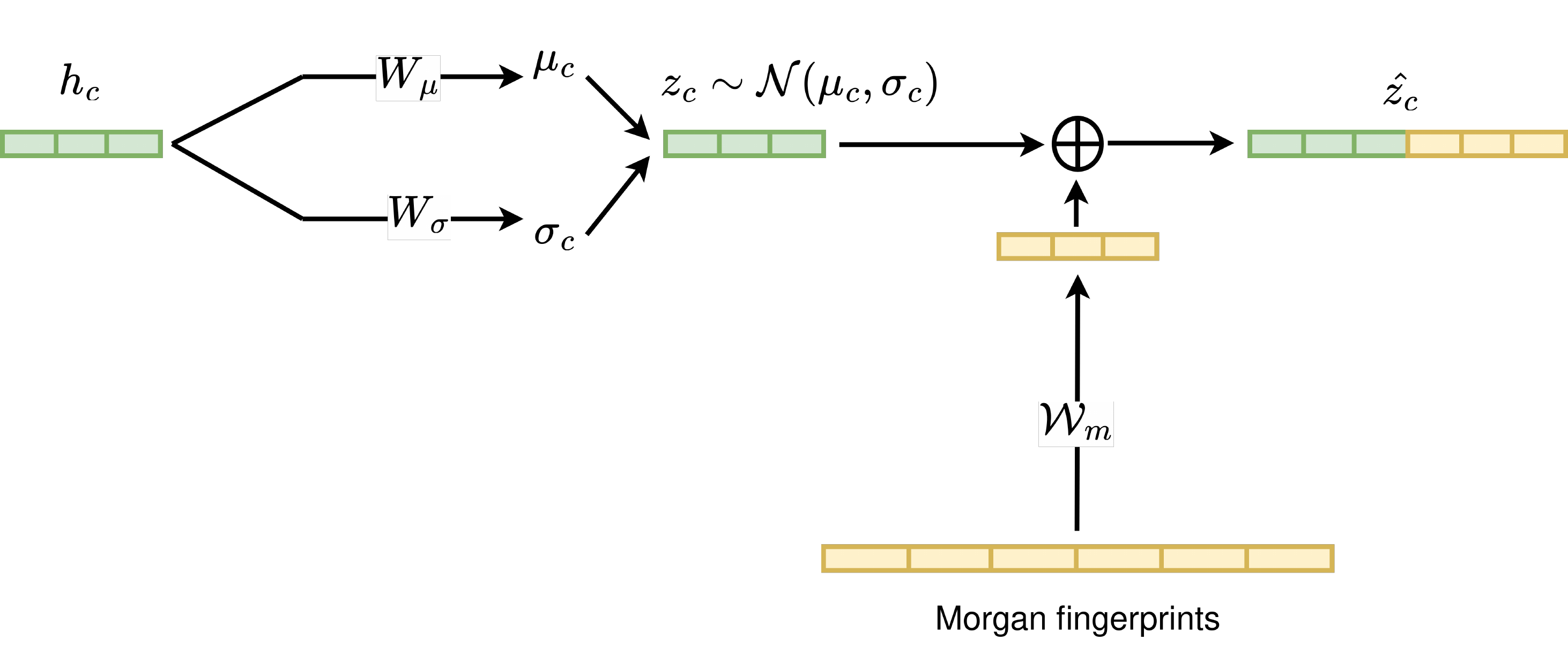}
    \caption{Overview of the latent encoder mentioned in \ref{latent_encoder}. Morgan fingerprints are concatenated with latent vector $z_c$ to improve the performance of polypharmacy side effects prediction. $\{W_\mu, W_\sigma, W_m$\} are trainable weight matrices.$\oplus$ denotes concatenation.} 
    \label{fig:3}
\end{figure}

\thickmuskip=0mu 
\paragraph{Latent Encoder} For each node type $v$, there are two multilayer perceptrons (MLPs) receiving the node embeddings from the encoder and computing the predicted mean $\mu$ and logarithm of the standard deviation $\log \sigma$ of the posterior latent distribution: 
    \[q_v (Z_v\mid X, E) = \prod_{i = 1}^{|V_v|}q_v(z^i_{v} \mid X, E)\]
    $q_v(z^i_v \mid X, E) = \mathcal{N}(z^i_v \mid \mu^i_v, \textrm{diag}((\sigma^i_v)^2)$ denotes the posterior distribution of a node of a specific node type. Here, $\mu_v$ and $\log \sigma _v$ are computed as follows:
    \[\mu_v = W^2_{\mu_v} \tanh(W^1_{\mu_v} h_v)\]
    \[\log \sigma_v = W^2_{\sigma_v} \tanh(W^1_{\sigma_v} h_v)\]
    where $W^i_{\mu_v} \in \mathbb{R} ^ {d_k \times d}$, $W^i_{\sigma_v} \in \mathbb{R} ^ {d_k \times d}$ are the weight matrices, $\mu_v$ and $\log \sigma_v$ are the matrices of mean vector $\mu^i_v$ and logarithm of standard deviation vector $\log \sigma^i_v$, respectively. Figure \ref{fig:3} demonstrates how we integrate molecular structures of drug nodes with their latent representation to improve the performance in side effect link prediction.

\paragraph{Decoder} a tensor factorization using latent embeddings to predict the node interactions on $G$. We follow the approach proposed in \cite{10.1093/bioinformatics/bty294} to design the decoder: 
    \begin{equation*}
        g(v_i, e, v_j) = 
        \begin{cases}
            z^T_i D_e R D_e z_j & \text{if $v_i$ and $v_j$ are drugs} \\
            z^T_i M_e z_j & \text{if $v_i$ and $v_j$ are a protein and a drug, or vice versa.}
        \end{cases}
    \end{equation*}
    where $D_e, R, M_e \in \mathbb{R} ^ {d \times d}$ are learnable parameters. $R$ denotes the global matrix representing all drug-drug interactions among all polypharmacy side effects; $M_e$ is a edge-type-specific matrix modeling drug-protein and protein-protein relations.
    Also, $D_e$ is a diagonal matrix, and its on-diagonal entries model the significance factors of $z_i$ and $z_j$ in multiple dimensions under the side effect type $e$. Finally, the probability of edge $(v_i, e, v_j)$ is calculated via a sigmoid function $\sigma$: 
    \[p_e(v_i, v_j) = \sigma(g(v_i, e, v_j))\]
  
\section{Experiments}
\label{sec:3}

Our implementation is done with the PyTorch Geometric library \cite{fey2019fast, paszke2019pytorch}. The code base is developed based on the implementation of VGAE published in the library's repository. We release our implementation at \url{https://github.com/HySonLab/drug-interactions}.

\subsection{Polypharmacy side effects}
\begin{table}[h]
	\centering
	\caption{Statistics for each type of interaction}
	\vskip 1em
	\label{tab:sample}
	\begin{tabular}{cc}
		\toprule
		Interaction & $\#$ Edges  \\
		\midrule
        Drug-Drug & 4,651,131 \\
        Drug-Protein & 18,596 \\
        Protein-Protein & 715,612 \\
		\bottomrule
	\end{tabular}
	\label{tab:1}
\end{table}
We conduct experiments on the dataset introduced in \cite{10.1093/bioinformatics/bty294}. It is a multimodal graph consisting of 19,085 protein and 645 drug nodes. There are three types of interaction including drug-drug, drug-protein, and protein-protein. In particular, the dataset contains 964 commonly occurring polypharmacy side effects, resulting in 964 edge types of drug-drug interaction. Protein-protein and drug-protein interactions are regarded as two other edge types; therefore, the multimodal graph has 966 edge types in total. Additionally, total numbers of edges of each interaction are displayed in Table \ref{tab:sample}.

We randomly split the edges into training, validation, and testing sets with a ratio of $8 : 1 : 1$. Then, the edges in the training edge set are randomly divided into $80\%$ edges for message passing and $20\%$ edges for supervision. We use Adam optimizer \cite{bock2018improvement} to minimize the loss function $L$ showed in Equation \ref{eq:1} in 300 epochs with a learning rate of $0.001$. We run the experiments with six different random seeds. 

\begin{equation}
    L = \sum_{(v_i, e, v_j) \in E}-\log p_e(v_i, v_j) - \mathbb{E}_{v_n \sim P_e(v_i)}[ 1 - \log p_e(v_i, v_n)] - \sum_{v} \lambda_v \mathbb{D}_{KL}(q_v(Z_v\mid X, E)\mid p_v(Z_v))
    \label{eq:1}
\end{equation}
where $p_v(Z_v)$ denotes the latent space's prior distribution of the node type $v$. The first term of $L$ denotes the cross-entropy loss of the probabilities of positive and negative edges which is sampled by choosing a random node $v_n$ for each node $v_i$ under a specific edge type $e$, while the second term is the weighted sum of KL-divergence of each node type $v$; $\{\lambda_v\}$ are hyper-parameters. In our experiments, $\lambda_d$ and $\lambda_p$ equal to $0.9$ and $0.9$ for drug and protein nodes, respectively. Finally, we use one-hot representations indicating node indices as features for both drug and protein nodes.

\subsection{Anticancer drug response}
In addition to drug-pair polypharmacy side effects, we evaluate our approach for the anticancer drug response prediction problem; the task is  to predict the interactions between drug and cell line on a multimodal network. 
Integrated information between drugs and cell lines are an effective approach to calculate anticancer drug responses using computational methods; various type of information can be exploited such as drug chemical structures, gene expression profiles, etc., to provide more accurate predictions. In this work, we evaluate the performance of VGAE on the CCLE dataset proposed in \cite{ahmadi2020adrml} which contains comprehensive information of drugs and cell lines. Three types of pairwise similarities are provided for each modality, resulting in nine types of combination used in computation among drug and cell line pairs; in addition, pairwise logarithms of the half-maximal inhibitory concentration (IC50) scores between drugs and cell lines are given as values to be predicted. The combinations are detailed in Table \ref{tab:3}.

To apply VGAE in predicting the IC50 scores among drug-cell line pairs, we construct an undirected multimodal graph consisting of drug and cell line nodes with the similarity scores indicating the edge weights under three types of relation, including drug-drug, cell line-cell line, and drug-cell line. The problem is regarded as a link prediction task in which an amount of edges connecting drug and cell line nodes are masked, and the objective is to compute real values $w_e$ denoting the weights of these missing links. The implementation of VGAE is almost the same as detailed in \ref{sec:sub2} with modifications of the decoding stage in which representations of two nodes are concatenated before being preceded to a simple multilayer perceptron with two hidden layers of size 16, ReLU nonlinearity, and a linear layer on top to predict a final score denoting the edge weights among them. 

We randomly split the edges with a ratio 7:1:2 for training, validation, and testing purposes, respectively. In this task, the models are trained to reconstruct a weighted adjacency matrix and regularized by KL-divergence of the node latent distributions with hyperparameters $\lambda_d$ and $\lambda_c$ denoting the coefficients for drug and cell nodes, respectively; $\lambda_d = \lambda_c = 0.001$ in our experiments. We train the models in 500 epochs with a learning rate of 0.01 to minimize the loss function $L$ as follows: 
\begin{equation}
    L = \sum_{(v_i, e, v_j) \in E} (\widehat{s_e}(v_i, v_j) - s_e(v_i, v_j))^2
    - \sum_{v} \lambda_v \mathbb{D}_{KL}(q_v(Z_v\mid X, E)\mid p_v(Z_v))
    \label{eq:2}
\end{equation}
where $\widehat{s_e}(v_i, v_j)$ indicates the predicted edge weights between node $v_i$ and $v_j$, whereas $s_e(v_i, v_j)$ are their ground truths. Finally, experiments are run with 20 different random seeds.

\section{Results} 
\subsection{Polypharmacy side effects prediction}
\begin{table}[h]
	\centering
	\caption{Average performance on polypharmacy link prediction across all side effects}
	\vskip 1.2em
	\label{tab:sample}
	\begin{tabular}{cccc}
		\toprule
		Method & AUROC & AUPRC & AP@50 \\
		\midrule
		RESCAL tensor factorization & 0.693 & 0.613 & 0.476 \\
		DEDICOM tensor factorization & 0.705 & 0.637 & 0.567 \\
		DeepWalk neural embeddings & 0.761 & 0.737 & 0.658 \\
		Concatenated drug features & 0.793 & 0.764 & 0.712 \\
	    Decagon & 0.872 & 0.832 & 0.803 \\
	    GAE & 0.893 $\pm$ 0.002 & 0.862 $\pm$ 0.003 & 0.819 $\pm$ 0.006 \\
	    \midrule
		VGAE (ours) & 0.905 $\pm$ 0.001 & 0.880 $\pm$ 0.001 & 0.853 $\pm$ 0.005 \\
		VGAE + Morgan fingerprints (ours) & \textbf{0.944 $\pm$ 0.005} & \textbf{0.926 $\pm$ 0.005} & \textbf{0.920 $\pm$ 0.004} \\
		\bottomrule
	\end{tabular}
	\label{tab:2}
\end{table}
We compare the performance of VGAE to alternative approaches. In addition to VGAE, we also implement a graph autoencoder and train the model in the same setting detailed in Sec. \ref{sec:3}. The baseline results are taken from \cite{10.1093/bioinformatics/bty294} including: RESCAL Tensor Factorization \cite{nickel2011three}, DEDICOM Tensor Factorization \cite{papalexakis2016tensors}, DeepWalk Neural Embeddings \cite{perozzi2014deepwalk, zong2017deep}, Concatenated Drug Features, and Decagon \cite{10.1093/bioinformatics/bty294}. It is worth noting that Decagon is also a GAE-based model with two GCN layers, yet \citet{10.1093/bioinformatics/bty294} use side information (e.g., side effects of individual drugs) as additional features for drug nodes. By constrast, our VGAE and GAE are trained on one-hot representations for drug and protein nodes. To examine the effects of Morgan fingerprints in drug-drug interaction decoding, we augment the Morgan fingerprint information into the latent vectors of drug nodes before preceding them to tensor-based decoders to produce final predictions.

Table \ref{tab:2} shows the results of all approaches in predicting the polypharmacy side effects. The models are evaluated based on three metrics, including area under ROC curve (AUROC), area under precision-recall curve (AUPRC), and average precision at 50 (AP@50). The scores reveal that VGAE without augmenting Morgan fingerprints outperforms traditional tensor-based approaches by a large margin, resulting in up to $24.8 \%$ (AUROC), $32.0 \%$ (AUPRC), and $40.3 \%$ (AP@50). We also compare VGAE with two machine-learning-based methods. The model achieves a $25.8 \%$ gain over DeepWalk neural embeddings and $18.0 \%$ over Concatenated drug features in AP@50 scores.
Furthermore, albeit trained on one-hot feature vectors, vanilla VGAE can achieve competitive performance with Decagon, outperforming the baseline by $6.0\%$ (AP@50). This indicates the effectiveness of VGAE in recommending potential side effects in a featureless drug-protein multi-modal network. Finally, VGAE with additional molecular information at the decoding stage provides the best performance across the approaches. The method achieves the highest scores among all three metrics and outperforms the others by a large margin. The results reveal that using molecular fingerprints is a simple yet effective solution in predicting drug-drug interactions as suggested in \cite{long500molecular}.
\begin{figure}[h]
    \centering
    \includegraphics[width = 0.7\textwidth]{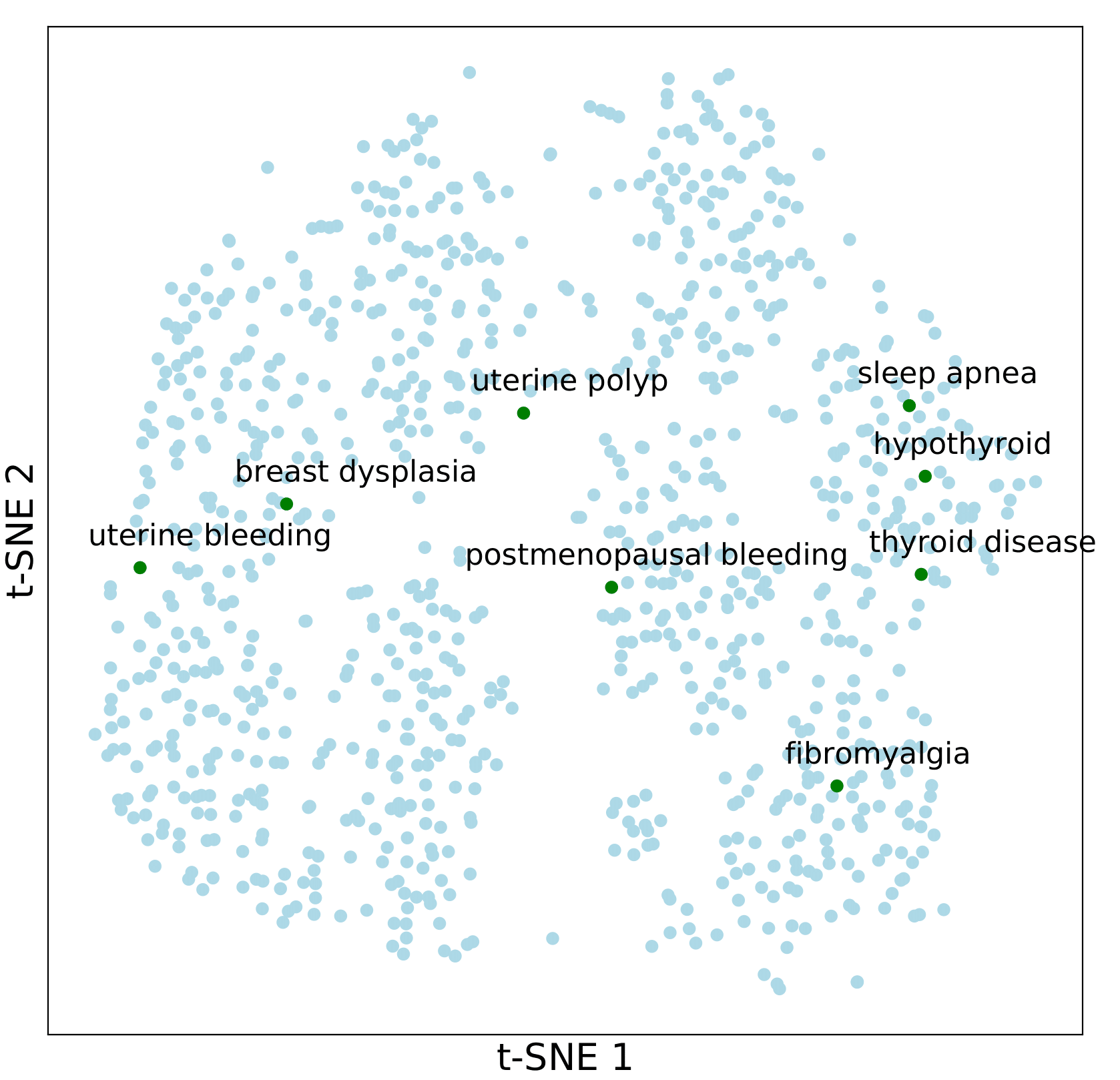}
    \caption{Visualization of polypharmacy side effects modeled by VGAE}
    \label{fig:1}
\end{figure}

\paragraph{Visualization of side effect embeddings}
To further demonstrate the capability of VGAE in learning to model drug-pair polypharmacy side effects, we plot their representations in 2D dimension using the t-SNE \cite{van2008visualizing} method. The embedding of each side effect is derived by taking on-diagonal entries of the tensor $D_e$ to create a $d$-dimensional vector which is accordingly projected into 2D space. Figure \ref{fig:1} illustrates a vector space in which  side effects' representations establish a clustering structure.

\subsection{Anticancer drug response prediction}
In this task, the models are evaluated using four criteria including root mean square error (RMSE), coefficient of determination ($R^2$), Pearson correlation coefficient (PCC), and fitness which is computed as: 
\[fitness \,=\, R^2 + PCC - RMSE\]

Table \ref{tab:3} shows the experimental results of drug response prediction on nine combinations of pairwise similarities of drugs and cell lines. VGAE trained on copy number variation with KEGG-pathways as similarity scores for cell lines and drugs achieve the best performance, yielding the lowest RMSE at 0.46 $\pm$ 0.02 and highest scores in $R^2$ = 0.67 $\pm$ 0.03, PCC = 0.85 $\pm$ 0.01, and fitness = 1.05 $\pm$ 0.06. Moreover, Table \ref{tab:4} demonstrates the comparisons between VGAE and other baselines which are taken from \cite{ahmadi2020adrml} including CDRscan
\cite{chang2018cancer}, CDCN \cite{wei2019comprehensive}, SRMF \cite{suphavilai2018predicting}, CaDRRes \cite{wang2017improved}, ADRML \cite{ahmadi2020adrml}, and
k-nearest neighbors (KNN). The results reveal that compared with the baselines, VGAE has competitive performance, especially in the fitness score determining the effectiveness of the approaches in the drug response prediction task. 
\begin{table}[h]
	\centering
	\caption{Performance of VGAE on various types of similarities of drugs and cell lines}
	\vskip 1em
	\label{tab:sample}
	\begin{tabular}{cccccc}
		\toprule
		Cell line similarity & Drug similarity & RMSE $\downarrow$ & $R^2$ $\uparrow$ & PCC $\uparrow$ & fitness $\uparrow$ \\
		\midrule
		
		Mutation & Chemical & 0.48 $\pm$ 0.02 & 0.64 $\pm$ 0.04 & 0.84 $\pm$ 0.01 & 1.00 $\pm$ 0.06 \\
		Gene expression & Chemical & 0.47 $\pm$ 0.01 & 0.66 $\pm$ 0.03 & 0.85 $\pm$ 0.00 & 1.03 $\pm$ 0.05 \\
		Copy number variation & Chemical & 0.47 $\pm$ 0.02 & 0.65 $\pm$ 0.03 & 0.84 $\pm$ 0.01 & 1.02 $\pm$ 0.05 \\
		Mutation & Target protein & 0.48 $\pm$ 0.02 & 0.66 $\pm$ 0.03 & 0.84 $\pm$ 0.01 & 1.01 $\pm$ 0.05 \\
		Gene expression & Target protein & 0.47 $\pm$ 0.02 & 0.66 $\pm$ 0.03 & 0.84 $\pm$ 0.01 & 1.03 $\pm$ 0.05 \\
		Copy number variation & Target protein & 0.48 $\pm$ 0.01 & 0.65 $\pm$ 0.03 & 0.84 $\pm$ 0.01 & 1.02 $\pm$ 0.05 \\
		Mutation & KEGG & 0.47 $\pm$ 0.02 & 0.66 $\pm$ 0.03 & 0.85 $\pm$ 0.01 & 1.03 $\pm$ 0.06 \\
		Gene expression & KEGG & 0.47 $\pm$ 0.02 & 0.65 $\pm$ 0.04 & 0.84 $\pm$ 0.01 & 1.03 $\pm$ 0.08 \\
		Copy number variation & KEGG & \textbf{0.46 $\pm$ 0.02} & \textbf{0.67 $\pm$ 0.03} & \textbf{0.85 $\pm$ 0.01} & \textbf{1.05 $\pm$ 0.06} \\
		\bottomrule
	\end{tabular}
	\label{tab:3}
\end{table}

\begin{table}[h]
	\centering
	\caption{Methods' performance in predicting anticancer drug response}
	\vskip 1em
	\label{tab:sample}
	\begin{tabular}{ccccc}
		\toprule
		Method  & RMSE $\downarrow$ & $R^2$ $\uparrow$ & PCC $\uparrow$ & fitness $\uparrow$ \\
		\midrule
		ADRML  & 0.49 & \textbf{0.68} & \textbf{0.85} & 1.04 \\
		CDRscan  & 0.76 & 0.67 & 0.83 & 0.74\\
		CDCN  & 0.48 & 0.67 & 0.83 & 1.02 \\
		SRMF  & \textbf{0.25} & 0.40 & 0.80 & 0.95 \\
	    CaDRRes & 0.53 & 0.31 & 0.52 & 0.3 \\
	    KNN  &  0.56 & 0.57 & 0.78 & 0.79 \\
	    \midrule 
	    VGAE (ours) & 0.46 $\pm$ 0.02 & 0.67 $\pm$ 0.03 & \textbf{0.85 $\pm$ 0.01} & \textbf{1.05 $\pm$ 0.06} \\
		\bottomrule
	\end{tabular}
	\label{tab:4}
\end{table}

\section{Conclusion}
In this work, we evaluate the effectiveness of variational graph autoencoders in predicting potential polypharmacy side effects on multimodal networks. The results reveal that VGAE trained on one-hot feature vectors outperforms other approaches. Moreover, augmenting Morgan fingerprints before the decoding stage helps boost the performance of VGAE. This suggests further examination of the use of molecular fingerprints in drug-drug interaction problems.
\bibliographystyle{unsrtnat}
\bibliography{reference}
\newpage
\appendix



\end{document}